# Mapping cellular magnesium using X-ray microfluorescence and atomic force microscopy

Stefano Lagomarsino[1], Stefano Iotti[2,7], Giovanna Farruggia[3], Alessia Cedola[1], Valentina Trapani[4], Michela Fratini[1], Inna Bukreeva[1,8], Andrea Notargiacomo[1], Lucia Mastrototaro[4], Ian McNulty[5], Stefan Vogt[5], Daniel Legnini[5], Sangsoo Kim[5], Jeanette A M Maier[6] & Federica I Wolf[4]

*1 IFN-CNR - V. Cineto Romano, 42 00156 Rome - Italy*
*2 Dipartimento di Medicina Interna, dell'Invecchiamento e Malattie Nefrologiche Università di Bologna, Via Massarenti, 9 40138 Bologna - Italy*
*3 Dipartimento di Biochimica « G. Moruzzi » Università di Bologna, Via Irnerio, 48 40126 Bologna - Italy*
*4 Istituto di Patologia Generale - Università Cattolica del Sacro Cuore - Facoltà di Medicina "A. Gemelli" L.go F. Vito, 1 00168 Rome – Italy*
*5 Argonne National Laboratory, 9700 South Cass Avenue, Argonne, Illinois 60439, USA*
*6 Dipartimento di Scienze Cliniche, Università di Milano, Via GB Grassi 74, 20157 Milan– Italy*
*7 Istituto Nazionale Biostrutture e Biosistemi - Rome – Italy*
*8 Lebedev Institute, Moscow - Russia*

**Abstract**

Magnesium is the most abundant intracellular divalent cation. We present an innovative experimental approach to localizing intracellular magnesium that combines elemental and morphological information from individual cells with high-resolution spatial information. Integration of information from scanning fluorescence X-ray microscopy with information from atomic force microscopy was used to generate a magnesium concentration map and to determine the X-ray linear absorption coefficient map within a whole dehydrated mammary epithelial cell.

Magnesium (Mg referring both to ionized and bound cation) plays crucial structural and regulatory roles within all cells, including stabilization of membrane bilayers, nucleic acids, and proteins[1], and modulation of enzymatic reactions such as transphosphorylations. In the face of the enormous amount of data about the biochemistry of Mg, a complete picture of its regulation and cellular homeostasis is lacking due both to conceptual difficulties and technical limitations.

Despite the abundance of Mg in biological samples and the lack of substantial concentration gradients between intra- and extracellular environments, regulation of intracellular magnesium is finely tuned[2]. Experimental evidence suggests that a regulatory system controls not only Mg influx/efflux through the cell membrane[3], but also Mg buffering and compartmentalization in intracellular organelles, particularly mitochondria[4]. Indeed, several reports indicate that mitochondrial Mg is mobilized by hormones and other stimuli[4,5]. These data suggest that the release of Mg from mitochondria might play a key role in cell survival or cell death[6,7], a pathway that has been advocated to be relevant in the genesis of the drug-resistance phenotype in tumour cells[8,9].

There is a need for investigative tools to map intracellular Mg in order to elucidate whether its regulatory mechanism stems from mobilization from intracellular stores and /or from the balance between free and bound Mg[2]. Despite recent efforts in applying new live imaging techniques to the field of magnesium research[10], an accurate characterization of Mg distribution in the cellular environment is still lacking. Here we used scanning fluorescence X-ray microscopy (SFXM), a highly sensitive method for mapping trace elements in cells[11]. Accurate mapping of intracellular Mg concentrations by SFXM can be performed on cell sections such as those utilized for electron microscopy. However, measuring a single section does not reveal the distribution of the target element throughout the whole cell. This study addressed the problem of mapping the Mg concentration, and not just the Mg content, in whole cells. Towards this end, we combined SFXM measurements with atomic force microscopy (AFM) measurements on the same cell while taking into account their non-homogeneous thickness. In this manner, we mapped the distribution of intracellular Mg in dehydrated whole cells and obtained a Mg concentration map.

SFXM is performed by scanning a sample in a micro-focused or nano-focused X-ray beam, as the fluorescence X-ray spectrum emitted by the sample is measured with an energy-dispersive detector. The fluorescence intensity measured at each scan point is proportional to the elemental content in the illuminated area. To determine the concentration of the target element, it is necessary to know the illuminated volume. Since the area is known and is identical for each scan point, the cell thickness must be measured and correlated to the SFXM signal at each point. This can be done by AFM in fixed/dehydrated cells provided that accurate spatial registration is carried out between the AFM and SFXM maps.

We utilized HC11 mammary epithelial cells since their total Mg content was previously characterized[12]. Cells were grown on 100-nm thick $Si_3N_4$ membranes and then rapidly dehydrated in methanol/acetone to minimize Mg loss that might occur with repeated washes. The SFXM measurements were carried out using the scanning X-ray microscope[13] at the 2-ID-B beamline at the Advanced Photon Source[14]. A Fresnel zone plate focused a monochromatic 1.5-keV beam to a spot size of about 50 nm on the sample. The sample was first examined with a phase contrast optical microscope to define the region of interest. The same area was then scanned in the X-ray beam as the Mg K fluorescence line was measured with a silicon drift diode detector and the X-ray intensity transmitted by the sample was simultaneously recorded with a 9-element configured detector that allowed for both transmitted intensity and differential phase contrast imaging[15] (see Fig. 1a and on-line method section).

Phase contrast X-ray microscopy is sensitive to sample thickness and/or density variation. However, obtaining quantitative 3-D information by this method requires detailed knowledge of the sample constituents and their complex refractive indices. Figure 1 shows the following images of a single HC11 cell: optical phase contrast, X-ray transmission, X-ray phase contrast, and Mg fluorescence (Fig. 1b-f). The fluorescence intensity has been normalized to the total Mg content of a HC11 cell (about 20 fmol, as measured by atomic absorption spectroscopy[12]), and the scale is expressed in femtograms.

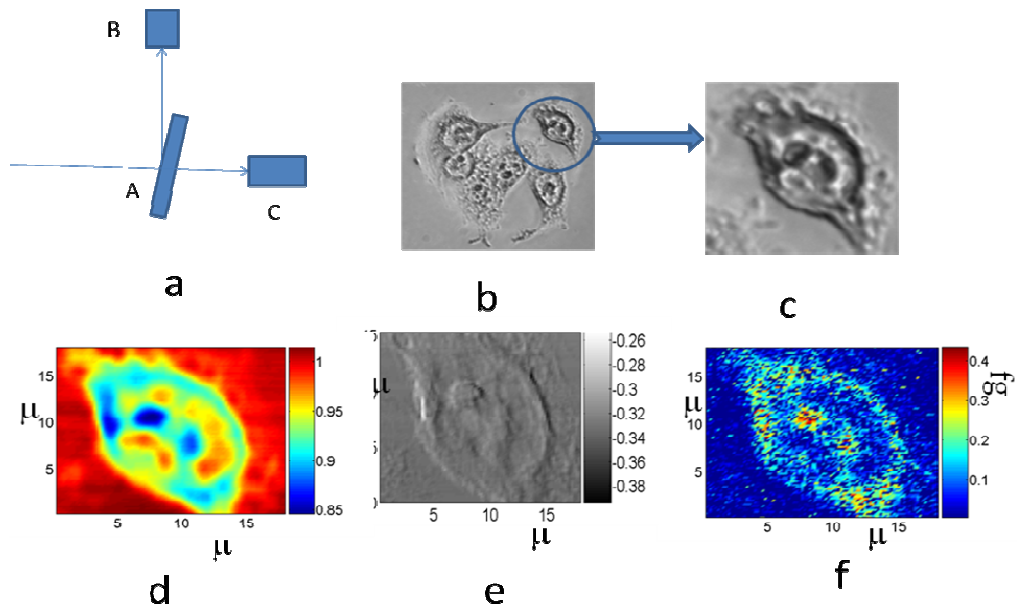

*Figure 1| (a) Relative orientation of the incident focused X-ray beam, sample, and fluorescence detector in the scanning X-ray microscope; A: sample; B: fluorescence detector; C: transmission detector. Optical phase contrast images of (b) a group of dehydrated HC11 cells and (c) a single cell measured by SFXM. (d) X-ray transmission and (e) X-ray phase-contrast images of the same cell, in arbitrary units. (f) the Mg fluorescence intensity map, expressed in femtograms.*

These images show that the dehydrated cell is nearly filled by a large thick nuclear region including some perinuclear area, surrounded by a thin cytoplasmic area. From the optical phase contrast image (Fig. 1b), it is also evident that it is difficult to define the cytoplasm borders of each cell. Note that the maximum fluorescence intensity corresponds to locations where there is maximum absorption (transmission image, Fig. 1d) and maximum thickness (phase contrast image, Fig. 1e). To obtain quantitative Mg concentration information, AFM was used to analyze the same cell studied by SFXM, and the fluorescence intensity was normalized using the 3-D thickness measurements obtained by AFM.

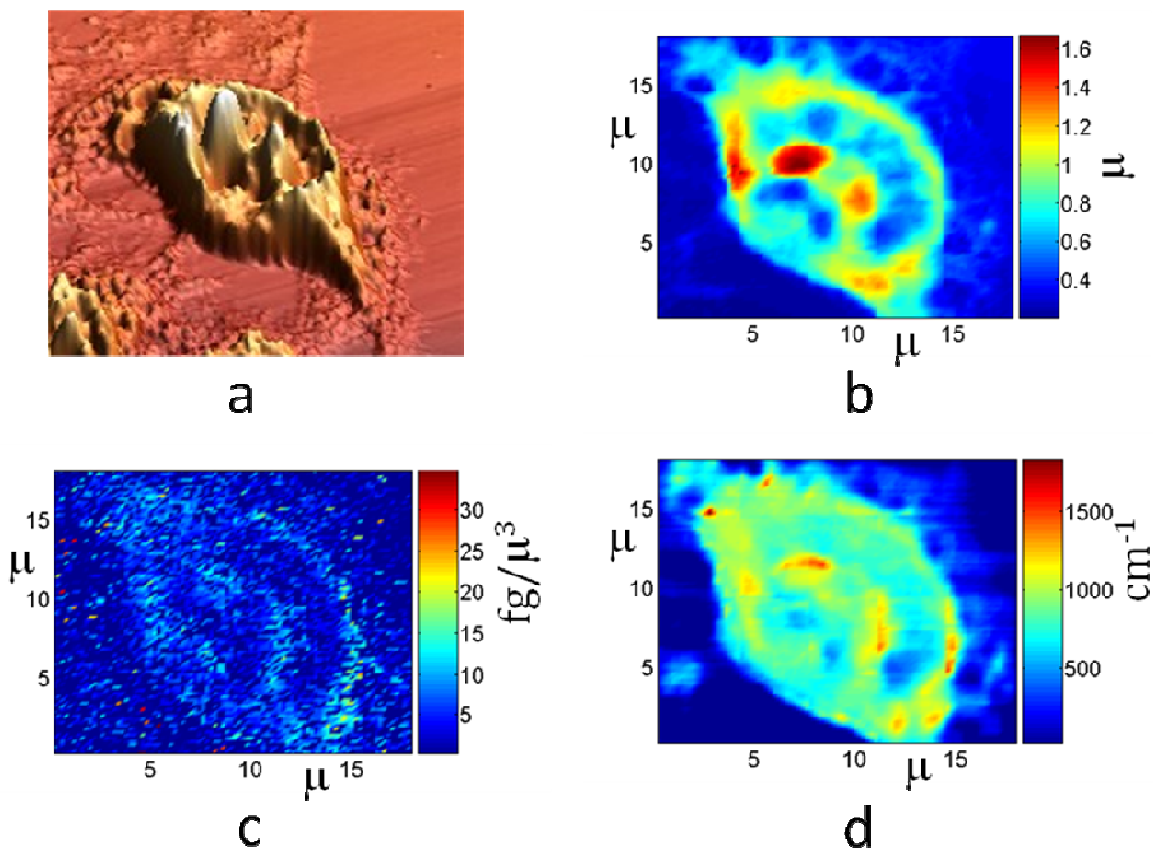

*Figure 2| (a) AFM 3-D rendering. (b) AFM thickness map, in microns. (c) Mg concentration map obtained by normalizing the fluorescence map with the measured cell thickness, in fg/$\mu^3$. (d) map of the X-ray linear absorption coefficient, $\tilde{\mu}$, in $cm^{-1}$. Abscissa and ordinates are in microns for all the figures.*

A 3-D rendering of the AFM data is shown in Figure 2a, and the thickness map is shown in Figure 2b. Using custom software written in MATLAB, we registered the fluorescence intensity map (shown in Fig. 1e) with the thickness map (Fig. 2b). The self-absorption of the fluorescent X-rays as they traversed the sample towards the detector was also taken into account, because this can give rise to shadowing effects. The correction was applied using the cell topography information obtained by AFM. Normalizing the fluorescence intensity obtained by SFXM, corrected by self-absorption, with the thickness measured by AFM, we obtained the concentration map shown in Figure 2c. In this case, the local concentration values were determined assuming the total Mg content in the cell, as measured by atomic absorption spectroscopy[12] (see above and on-line methods section). In future experiments, the use of a standard could provide a direct concentration measurement. The concentration map is different from the fluorescence intensity map shown in Figure 1e: in the nuclear region, the concentration is quite uniform, indicating that the strong intensity peaks found in the fluorescence map were only related to larger thickness (see Fig. 2a-b). Interestingly, in the perinuclear region, where most subcellular organelles are found, some areas with higher Mg concentration are noticeable that were not evident in the fluorescence intensity map.

Using the transmitted X-ray intensity, T, and cell thickness, t, we retrieved a map of the X-ray linear absorption coefficient, $\mu$ (Fig. 2d); $\mu$ is related to the imaginary part of the refraction index through the simple equation: $T = \exp(-\mu t)$. To our knowledge, this is the first time that such a map has been obtained for a single cell. The incident photon energy (1.5 keV) is ideal for this purpose, because at this energy the absorption from thicker cellular structures is on the order of 20-30% (see Fig. 1d).

In conclusion, we demonstrated that a combination of complementary experimental methodologies, *i.e.* SFXM, AFM, and phase contrast X-ray microscopy, enables determination of a concentration map of intracellular Mg. These techniques also allowed us to determine a map of the X-ray linear absorption coefficient, which can give important information about local electronic density. In regards to sensitivity, we can detect Mg in the $10^{-17}$ g range within a single pixel. The present measurements were carried out with 200-nm scan steps, whereas the beam size was on the order of 50 nm. Higher spatial resolution is attainable in principle, but there is a trade-off between spatial resolution, measuring time, scanned area, and radiation damage. In the example shown here, the scanned area was 20 x 20 mm$^2$, and the measurement time was less than 2 hours. Importantly, these data were obtained from a whole cell, rather than from sections (as with electron microscopy techniques). In this work we used methanol/acetone dehydrated samples, the next step will be to perform experiments in cells in a hydrated state, *i.e.* using a cryogenic approach. In addition, the procedure for registering the X-ray and AFM measurements can be greatly facilitated, for example, by nano-patterning the sample substrate. It is worth noting that the absorption coefficient map can also be obtained by transmission X-ray microscopy using a laboratory source which does not require a synchrotron source as scanning fluorescence microscopy.

The current technological challenge is to combine elemental and morphological imaging information from single cells with high-resolution spatial information in order to perform quantitative analysis of chemical species inside subcellular compartments. The approach described here has great potential as it provides a high-resolution map of quantified element distribution within a cell; such quantitative information complements that offered by live single cell imaging techniques. Finally, this novel methodology provides a foundation for interesting work such as measuring electron density modifications in cells in different physiological and pathological conditions.

**Acknowledgements**

The use of the Advanced Photon Source was supported by the U.S. Department of Energy, Office of Science, and Office of Basic Energy Sciences, Contract No. DE-AC02-06CH11357. Financial support was obtained from Italian Ministry for Education and Research, project PRIN 2007ZT39FN_003.